\documentclass[10pt]{article}
\usepackage{jheppub}
\usepackage{empheq}
\usepackage{amsmath,amsfonts,dsfont}
\usepackage{subcaption}
\allowdisplaybreaks
\DeclareMathOperator*{\R}{\mathrm{\partial_{\rho}\Theta}}
\DeclareMathOperator*{\T}{\mathrm{\partial_{\tau}\Theta}}
\DeclareMathOperator*{\A}{\mathrm{\partial_{\alpha}\Theta}}
\DeclareMathOperator*{\Ps}{\mathrm{\partial_{\psi}\Theta}}
\DeclareMathOperator*{\Frt}{\mathit{F_{\rho\tau}}}
\DeclareMathOperator*{\Fra}{\mathit{F_{\rho\alpha}}}
\DeclareMathOperator*{\Frp}{\mathit{F_{\rho\psi}}}
\DeclareMathOperator*{\Fta}{\mathit{F_{\tau\alpha}}}
\DeclareMathOperator*{\Ftp}{\mathit{F_{\tau\psi}}}
\DeclareMathOperator*{\Fap}{\mathit{F_{\alpha\psi}}}
\begin{document}
\title{On the D5-brane description of $\frac{1}{4}$-BPS Wilson loops in $\mathcal{N}=4$ super Yang-Mills theory}
\author[a]{Alberto Faraggi}
\author[a]{and Crist\'obal Moreno}
\emailAdd{afaraggi@uc.cl}
\emailAdd{cjmoreno@uc.cl}
\affiliation[a]{Instituto de F\'isica, Pontificia Universidad Cat\'olica de Chile,\\
Av. Vicu\~na Mackenna 4860, Santiago, Chile.}
\abstract{
We construct the probe D5-brane solution in $AdS_5\times S^5$ dual to the $\frac{1}{4}$-BPS latitude Wilson loop in $\mathcal{N}=4$ super Yang-Mills theory in the $k$-antisymmetric representation of $SU(N)$. The solution is exact in the latitude parameter $\theta_0$ and correctly reproduces the $\frac{1}{2}$-BPS limit. We compute the string charge $k$ and the renormalized on-shell action perturbatively to order $\mathcal{O}\left(\theta_0^{10}\right)$ and find full agreement with the expectation value of the Wilson loop predicted by the Gaussian matrix model in the limit $N\sim k\to\infty$, $\lambda\to\infty$.
}
\maketitle
\section{Introduction}
Since the early works of \cite{Rey:1998ik,Maldacena:1998im}, Wilson loop operators have played a central role in the development of the AdS/CFT correspondence \cite{Maldacena:1997re}. In $\mathcal{N}=4$ super Yang-Mills theory they are defined as
\begin{empheq}{alignat=7}
	W_{\mathcal{R}}[\mathcal{C}]&=\textrm{Tr}_{\mathcal{R}}\mathcal{P}\exp\left(\oint_{\mathcal{C}}d\tau\left(A_{\mu}\dot{x}^{\mu}+|\dot{x}|n_I\phi^I\right)\right)\,,
\end{empheq}
where $\mathcal{C}$ labels a closed curve parametrized by $x^{\mu}(\tau)\in\mathds{R}^4$ and $n_I(\tau)\in\mathds{R}^6$, $\mathcal{R}$ is a representation of the $SU(N)$ gauge group, and the symbol $\mathcal{P}$ denotes path ordering along the loop. Of particular interest to us is the one-parameter family of operators given by \cite{Drukker:2005cu,Drukker:2006ga}
\begin{empheq}{alignat=7}\label{eq: latitude couplings}
	x^{\mu}(\tau)&=\left(\cos\tau,\sin\tau,0,0\right)\,,
	&\qquad
	n_I(\tau)&=\left(0,0,0,\sin\theta_0\cos\tau,\sin\theta_0\sin\tau,\cos\theta_0\right)\,,
\end{empheq}
with
\begin{empheq}{alignat=7}
	0&\leq\theta_0&&\leq\frac{\pi}{2}\,.
\end{empheq}
These so-called latitude Wilson loops preserve a $U(1)\times SO(3)\times SO(3)\subset SO(2,4)\times SO(6)$ bosonic symmetry as well as $8$ of the $32$ supercharges of $\mathcal{N}=4$ super Yang-Mills, thus forming the supergroup $SU(2|2)\subset SU(2,2|4)$. In the $\theta_0\to0$ limit the symmetries are enhanced to $SL(2;\mathds{R})\times SO(3)\times SO(5)\subset OSp(4^*|4)$ and we recover the well-known $\frac{1}{2}$-BPS circular Wilson loop \cite{Erickson:2000af,Drukker:2000rr}. At the other end of the interpolation we get a special case of the Zarembo loops constructed in \cite{Zarembo:2002an}. For related work see \cite{Drukker:2007dw,Drukker:2007qr,Drukker:2007yx}.

Latitude Wilson loops have served as fertile ground for precision tests of AdS/CFT \cite{Forini:2015bgo,Faraggi:2016ekd,Aguilera-Damia:2018twq,Forini:2017whz,Cagnazzo:2017sny,Medina-Rincon:2018wjs}, mainly due to the existence of exact results. The expectation value of these operators was conjectured \cite{Drukker:2006ga,Drukker:2007qr,Drukker:2007yx} to be the same as that of the circular loop with the proviso that
\begin{empheq}{alignat=7}
	\lambda&\to\lambda'&&=\lambda\cos^2\theta_0\,,
\end{empheq}
where $\lambda=g_{YM}^2N$ is the 't Hooft coupling. This was later proven using localization in \cite{Pestun:2007rz,Pestun:2009nn}. For the fundamental representation of $SU(N)$ the Gaussian matrix model yields \cite{Erickson:2000af,Drukker:2000rr}
\begin{empheq}{alignat=7}\label{eq: vev F}
	\langle W_{\Box}(\theta_0)\rangle&=\frac{1}{N}L^1_{N-1}\left(-\frac{\lambda'}{4N}\right)e^{\frac{\lambda'}{8N}}&&\underset{N\to\infty}{\longrightarrow}\frac{2}{\sqrt{\lambda'}}I_1\left(\sqrt{\lambda'}\right)&&\underset{\lambda\to\infty}{\longrightarrow}e^{\sqrt{\lambda'}}\,.
\end{empheq}
In the $k$-symmetric representation the result is \cite{Hartnoll:2006is}
\begin{empheq}{alignat=7}\label{eq: vev S}
	\langle W_{\mathcal{S}_k}(\theta_0)\rangle&\underset{N,k,\lambda\to\infty}{\longrightarrow}e^{{\textstyle2N\left(\kappa'\sqrt{1+\kappa'^2}+\textrm{arcsinh}\,\kappa'\right)}}\,,
	&\qquad
	\kappa'&=\frac{k\sqrt{\lambda'}}{4N}\,,
\end{empheq}
whereas for the $k$-antisymmetric representation of $SU(N)$ the expectation value reads \cite{Hartnoll:2006is,Yamaguchi:2006tq}
\begin{empheq}{alignat=7}\label{eq: vev A}
	\langle W_{\mathcal{A}_k}(\theta_0)\rangle&\underset{N,k,\lambda\to\infty}{\longrightarrow}e^{{\textstyle\frac{2N\sqrt{\lambda'}}{3\pi}\sin^3\theta_k}}\,,
	&\qquad
	\frac{k\pi}{N}&=\theta_k-\cos\theta_k\sin\theta_k\,.
\end{empheq}
The limits are taken in the order $N\to\infty$ and $k\to\infty$ with $k/N$ fixed, and then $\lambda\to\infty$, with $\kappa$ fixed in the case of $\mathcal{S}_k$. The representation $\mathcal{A}_k$ is defined for $k\leq N$ and exhibits the symmetry $k\to N-k$.

At strong coupling Wilson loops have a holographic description in terms of macroscopic strings and D-branes. The dictionary was spelled out in \cite{Gomis:2006sb,Gomis:2006im} and states that a Wilson loop in the fundamental representation of $SU(N)$ is dual to a fundamental string in $AdS_5\times S^5$, whereas the $k$-symmetric and $k$-antisymmetric representations at $k\sim N$ are captured by probe D3- and  D5-branes, respectively, carrying $k$ units of string charge. For larger representations of rank $k\sim N^2$ the gravitational description is realized in terms of fully back-reacted bubbling geometries \cite{Yamaguchi:2006te,Lunin:2006xr,DHoker:2007mci}. The F1 and D3-brane solutions dual to the $\frac{1}{4}$-BPS latitude Wilson loops appeared in the literature long ago \cite{Drukker:2006ga,Drukker:2006zk}. However, to the best of our knowledge, the analogous D5-brane configuration is yet to be found. Our goal in this note is to construct such solution.

The paper is organized as follows. In section \ref{sec: background} we review the $AdS_5\times S^5$ background in suitable coordinates. Section \ref{sec: solution} is devoted to writing an appropriate ansatz for the D5-brane and then finding and solving the BPS equations. In section \ref{sec: on-shell} we compute the string charge and on-shell action. We conclude in section \ref{sec: conclusions} with a brief discussion of our results.
\section{Supergravity background}\label{sec: background}
Let us begin by reviewing the background supergravity fields. We work in Euclidean signature (see appendix \ref{app: conventions} for our notation and conventions). The target space metric is that of $AdS_5\times S^5$ with equal radii, namely,
\begin{empheq}{alignat=7}\label{eq: background metric}
	ds^2&=L^2\left(ds^2_{AdS_5}+d\Omega_5^2\right)\,,
	&\qquad
	L&=\left(4\pi g_s\alpha'^2N\right)^{\frac{1}{4}}\,.
\end{empheq}
It is supported by the self-dual Ramond-Ramond field strength
\begin{empheq}{alignat=7}\label{eq: background RR}
	F_{(5)}&=4L^4\left(-i\,\textrm{vol}\left(AdS_5\right)+\textrm{vol}\left(S^5\right)\right)\,,
	&\qquad
	*F_{(5)}&=-iF_{(5)}\,.
\end{empheq}
Here $\alpha'$ is the string slope parameter, $g_s$ is the string coupling constant and $N$ is the number of background D3-branes which source the $5$-form flux. Indeed,
\begin{empheq}{alignat=7}
	\frac{i}{2\kappa_{10}^2}\oint_{S^5}*F_{(5)}&=NT_{D3}\,,
	&\qquad
	2\kappa_{10}^2&=(2\pi)^7\alpha'^4g_s^2\,,
	&\qquad
	T_{D3}&=\frac{1}{(2\pi)^3\alpha'^2g_s}\,,
\end{empheq}
where $\kappa_{10}$ is related to the $10$-dimensional Newton's constant and $T_{D3}$ is the D3-brane tension (charge). The AdS/CFT dictionary identifies $N$ with the rank of the $SU(N)$ gauge group and $4\pi g_s=g_{YM}^2$. Equivalently, $L^2=\alpha'\sqrt{\lambda}$.

It is convenient to write the $AdS_5$ metric as a foliation over $AdS_2\times S^2$, that is,
\begin{empheq}{alignat=7}\label{eq: AdS5 metric}
	ds^2_{AdS_5}&=du^2+\cosh^2u\left(d\rho^2+\sinh^2\rho\,d\tau^2\right)+\sinh^2u\left(d\eta^2+\sin^2\eta\,d\xi^2\right)\,,
\end{empheq}
with
\begin{empheq}{alignat=7}\label{eq: AdS5 coordinate range}
    u&\geq0\,,
	&\qquad
	\rho&\geq0\,,
	&\qquad
	\tau&\sim\tau+2\pi\,,
	&\qquad
	0&\leq\eta\leq\pi\,,
	&\qquad
	\xi&\sim\xi+2\pi\,.
\end{empheq}
This makes the $SO(1,2)\times SO(3)\subset SO(1,5)$ isometries manifest. For the $5$-sphere we use
\begin{empheq}{alignat=7}\label{eq: S5 metric}
	d\Omega_5^2&=d\Theta^2+\sin^2\Theta\left(d\alpha^2+\cos^2\alpha\,d\psi^2+\sin^2\alpha\left(d\vartheta^2+\sin^2\vartheta\,d\varphi^2\right)\right)\,,
\end{empheq}
with
\begin{empheq}{alignat=7}\label{eq: S5 coordinate range}
	0&\leq\Theta\leq\pi\,,
	&\qquad
	0&\leq\alpha\leq\frac{\pi}{2}
	&\qquad
	\psi&\sim\psi+2\pi\,,
	&\qquad
	0&\leq\vartheta\leq\pi\,,
	&\qquad
	\varphi&\sim\varphi+2\pi\,.
\end{empheq}
This is the usual foliation of $S^5$ over $S^4$, except that the $4$-sphere is written as a foliation over $S^1\times S^2$. Here the $U(1)\times SO(3)\subset SO(6)$ isometries are manifest. The embedding coordinates $\vec{X}\in\mathds{R}^{1,5}$ and $\vec{Y}\in\mathds{R}^6$ that give rise to \eqref{eq: AdS5 metric} and \eqref{eq: S5 metric} are
\begin{empheq}{alignat=7}\label{eq: embedding}
	\vec{X}&=\left(
	\begin{array}{c}
		\cosh u\cosh\rho
		\\
		\cosh u\sinh\rho\cos\psi
		\\ 
		\cosh u\sinh\rho\sin\psi
		\\
		\sinh u\sin\eta\cos\xi
		\\
		\sinh u\sin\eta\cos\xi
		\\
		\sinh u\cos\eta
	\end{array}
	\right)\,,
	&\qquad
	\vec{Y}&=\left(
	\begin{array}{c}
		\sin\Theta\sin\alpha\sin\vartheta\sin\varphi
		\\
		\sin\Theta\sin\alpha\sin\vartheta\cos\varphi
		\\
		\sin\Theta\sin\alpha\cos\vartheta
		\\
		\sin\Theta\cos\alpha\cos\psi
		\\
		\sin\Theta\cos\alpha\sin\psi
		\\
		\cos\Theta
	\end{array}
	\right)\,.
\end{empheq}
Finally, the $4$-form potential reads
\begin{empheq}{alignat=7}\label{eq: RR 4-form}
	C_{(4)}&=4L^4\Big(-if_1(u)\textrm{vol}\left(AdS_2\times S^2\right)+f_2(\Theta)\textrm{vol}\left(S^4\right)\Big)\,,
\end{empheq}
where
\begin{empheq}{alignat=7}\label{eq: RR 4-form functions}
	f_1(u)&=-\frac{u}{8}+\frac{1}{32}\sinh(4u)\,,
	&\qquad
	f_2(\Theta)&=\frac{3\Theta}{8}-\frac{1}{4}\sin(2\Theta)+\frac{1}{32}\sin(4\Theta)\,.
\end{empheq}
We set $L=1$ henceforth.
\section{D5-brane solution}\label{sec: solution}
The dynamics of a probe D5-brane in $AdS_5\times S^5$ is governed by the action
\begin{empheq}{alignat=7}\label{eq: D5 action}
	S_{\textrm{D5}}&=T_{D5}\int d^6\sigma\sqrt{\det\left(g+2\pi\alpha'F\right)}-iT_{D5}\int 2\pi\alpha'F\wedge P[C_{(4)}]\,,
\end{empheq}
where $\sigma^a$, $a=1,\ldots,6$ are worldvolume coordinates, $P[\;]$ denotes the pullback from the target space to the worldvolume, $g_{ab}=P[G]_{ab}$ is the induced metric on the brane, and $F_{ab}=\partial_aA_b-\partial_bA_a$ is the field strength of the worldvolume  gauge field. The D5-brane tension is
\begin{empheq}{alignat=7}
	T_{D5}&=\frac{1}{(2\pi)^5\alpha'^3g_s}&&=\frac{N\sqrt{\lambda}}{8\pi^4}\,.
\end{empheq}
From now on we will absorb the factor of $2\pi\alpha'$ in the definition of $F_{ab}$.
\subsection{Ansatz}
We will work in a static gauge where the worldvolume coordinates are $\sigma^a=\left(\rho,\tau,\alpha,\psi,\vartheta,\varphi\right)$. As required by the holographic dictionary, this choice implies that the D5-brane pinches the circle parametrized by $\tau$ at the boundary $\rho\to\infty$ of $AdS_5$. The most general electric ansatz consistent with the $U(1)\times SO(3)\times SO(3)$ symmetries of the $\frac{1}{4}$-BPS latitude Wilson loop is
\begin{empheq}{alignat=7}\label{eq: D5 anstaz}
	u&=0\,,
	&\qquad
	\Theta&=\Theta(\rho,\alpha,\Delta)\,,
	&\qquad
	F_{ab}&=F_{ab}(\rho,\alpha,\Delta)\,,
	&\qquad
	F_{a\vartheta}&=0\,,
	&\qquad
	F_{a\varphi}&=0\,,
\end{empheq}
with
\begin{empheq}{alignat=7}\label{eq: definition of Delta}
	\Delta&\equiv\tau-\psi\,.
\end{empheq}
Indeed, the $S^2\subset AdS_5$ collapses at the origin of the base space, thus preserving the full $SO(3)$ symmetry of the sphere. An additional $SO(3)$ factor arises from the fact that nothing depends on the coordinates $(\vartheta,\varphi)$, so the worldvolume geometry inherits the isometry of $S^2\subset S^4$. This also requires turning off the gauge field components along $d\vartheta$ and $d\varphi$; a term proportional to $\sin\vartheta\,d\vartheta\wedge d\varphi$ is allowed by the $SO(3)$ symmetry but would source a magnetic charge. Finally, since the fields depend on $\tau$ and $\psi$ only through the difference $\Delta$, the $U(1)$ invariance is realized by a simultaneous shift of both angles.\footnote{As we will see below, this dependence is also required by supersymmetry.} Recall that the $\frac{1}{2}$-BPS solution \cite{Yamaguchi:2006tq} corresponds to
\begin{empheq}{alignat=7}\label{eq: 1/2-BPS solution}
	\Theta&=\theta_k\,,
	&\qquad
	F&=-i\cos\theta_k\sinh\rho\,d\rho\wedge d\tau\,,
\end{empheq}
where $0\leq\theta_k\leq\pi$ is a constant related to the electric charge $k$ that sources the gauge field by
\begin{empheq}{alignat=7}\label{eq: string charge 1/2 BPS}
	k&=\frac{N}{\pi}\left(\theta_k-\cos\theta_k\sin\theta_k\right)\,.
\end{empheq}
In this case the worldvolume geometry is $AdS_2\times S^4$. We should recover this in the $\theta_0\to0$ limit. 

The ansatz described above is too general to be useful so we will make some simplifying assumptions. Following the AdS/CFT dictionary, we consider the most straightforward extension of the vector of scalar couplings \eqref{eq: latitude couplings} into the bulk, namely,
\begin{empheq}{alignat=7}\label{eq: axis N}
	\vec{N}&=\left(
	\begin{array}{c}
		0
		\\
		0
		\\
		0
		\\
		\sin\theta\cos\tau
		\\
		\sin\theta\sin\tau
		\\
		\cos\theta
	\end{array}
    \right)\,,
\end{empheq}
where $\theta=\theta(\rho)$ is an undetermined function such that
\begin{empheq}{alignat=7}\label{eq: boundary condition theta}
    \theta&\underset{\rho\to\infty}{\longrightarrow}\theta_0\,.
\end{empheq}
In terms of the embedding coordinates \eqref{eq: embedding}, and motivated by the $\frac{1}{2}$-BPS solution \eqref{eq: 1/2-BPS solution}, we will look for configurations satisfying\footnote{Admittedly, this condition is not obvious a priori. It can be arrived at from the general ansatz by looking at the supersymmetry equations, solving for $\partial_a\Theta$, and demanding that the integrability conditions $\partial_a\partial_b\Theta=\partial_b\partial_a\Theta$ be satisfied. Here we have chosen to impose it from the beginning in order to simplify the presentation of the paper.}
\begin{empheq}{alignat=7}\label{eq: refined anstaz}
    \vec{Y}\cdot\vec{N}&=\cos\theta_k\,.
\end{empheq}
In other words, the D5-brane wraps an $S^4\subset S^5$ given by a constant latitude angle measured with respect to the axis $\vec{N}$, which itself depends on $\rho$ and $\tau$. The precise relation between the angle $\theta_k$ and the string charge $k$ must be determined via Gauss's Law, although we anticipate that \eqref{eq: string charge 1/2 BPS} remains true even for $\theta_0\neq0$. 

Condition \eqref{eq: refined anstaz} translates into an implicit equation for $\Theta(\rho,\alpha,\Delta)$, namely,
\begin{empheq}{alignat=7}\label{eq: beta=cos theta_k}
    \cos\theta\cos\Theta+\sin\theta\sin\Theta\cos\alpha\cos\Delta&=\cos\theta_k\,.
\end{empheq}
From here we can compute the derivatives
\begin{empheq}{alignat=7}
    \label{eq: R}
	\partial_{\rho}\Theta&=-\frac{\sin\theta\cos\Theta-\cos\theta\sin\Theta\cos\alpha\cos\Delta}{\cos\theta\sin\Theta-\sin\theta\cos\Theta\cos\alpha\cos\Delta}\,\partial_{\rho}\theta\,,
	\\
    \label{eq: A}
	\partial_{\alpha}\Theta&=-\frac{\sin\theta\sin\Theta\sin\alpha\cos\Delta}{\cos\theta\sin\Theta-\sin\theta\cos\Theta\cos\alpha\cos\Delta}\,,
	\\
    \label{eq: T}
	\partial_{\Delta}\Theta&=-\frac{\sin\theta\sin\Theta\cos\alpha\sin\Delta}{\cos\theta\sin\Theta-\sin\theta\cos\Theta\cos\alpha\cos\Delta}\,.
\end{empheq}
For the gauge field we adopt a potential of the form
\begin{empheq}{alignat=7}\label{eq: refined anstaz gauge field}
    A&=A_{\rho}d\rho+A_{\tau}d\tau\,.
\end{empheq}
In particular, this sets $F_{\alpha\psi}=0$. Unlike the D3-brane case, the radial component $A_{\rho}$ cannot be gauged away because it depends on the worldvolume coordinates $\alpha$ and  $\Delta$. Notice that this restricted ansatz is still invariant under the $U(1)\times SO(3)\times SO(3)$ symmetry.

Since the embedding of the D5-brane in $AdS_5\times S^5$ is determined by a single function $\Theta(\rho,\alpha,\Delta)$, the induced metric on the worldvolume can be written as
\begin{empheq}{alignat=7}\label{eq: induced metric}
	g_{ab}&=h_{ab}+\partial_a\Theta\partial_b\Theta\,,
\end{empheq}
with
\begin{empheq}{alignat=7}\label{eq: h metric}
	h_{ab}&\equiv\left(
	\begin{array}{cccc|cc}
		1 & 0 & 0 & 0 & 0 & 0
		\\
		0 & \sinh^2\rho & 0 & 0 & 0 & 0
		\\
		0 & 0 & \sin^2\Theta & 0 & 0 & 0
		\\
		0 & 0 & 0 & \sin^2\Theta\cos^2\alpha & 0 & 0
		\\
		\hline
		0 & 0 & 0 & 0 & \sin^2\Theta\sin^2\alpha & 0
		\\
		0 & 0 & 0 & 0 & 0 & \sin^2\Theta\sin^2\alpha\sin^2\vartheta
	\end{array}
	\right)\,.
\end{empheq}
From now on indices will be raised and lowered with the metric $h_{ab}$ and its inverse $h^{ab}$. After some algebra the expansion of the determinant in the Dirac-Born-Infeld action becomes\footnote{
The matrix $M^a_{\phantom{a}b}=\partial^a\Theta\partial_b\Theta+F^a_{\phantom{a}b}$ is effectively $4\times4$. One then has
\begin{empheq}{alignat*=7}
	\det\left(\mathds{1}+M\right)&=1+\textrm{Tr}\,M+\frac{1}{2!}\left(\textrm{Tr}^2M-\textrm{Tr}\,M^2\right)+\frac{1}{3!}\left(\textrm{Tr}^3M-3\,\textrm{Tr}\,M\,\textrm{Tr}\,M^2+2\,\textrm{Tr}\,M^3\right)
	\cr
	&+\frac{1}{4!}\left(\textrm{Tr}^4M-6\,\textrm{Tr}^2M\,\textrm{Tr}\,M^2+8\,\textrm{Tr}\,M\,\textrm{Tr}\,M^3+3\,\textrm{Tr}^2M^2-6\,\textrm{Tr}\,M^4\right)\,.
\end{empheq}
}
\begin{empheq}{alignat=7}
	\det\left(g+F\right)&=\det\left(h\right)\mathcal{L}\,,
\end{empheq}
where
\begin{empheq}{alignat=7}\label{eq: DBI}
	\mathcal{L}&=1+||d\Theta||^2+||F||^2+||d\Theta\wedge F||^2+\frac{1}{2!}||F\wedge F||^2\,.
\end{empheq}
Here we have abbreviated
\begin{empheq}{alignat=7}
	w&=\frac{1}{p!}w_{a_1\ldots a_p}d\sigma^{a_1}\wedge\cdots\wedge d\sigma^{a_p}\,,
	&\qquad
	||w||^2&=\frac{1}{p!}w^{a_1\ldots a_p}w_{a_1\ldots a_p}\,.
\end{empheq}
The explicit form of the DBI Lagrangian \eqref{eq: DBI} can be found in appendix \ref{app: explicit expressions}. On the other hand, the Wess-Zumino term reads
\begin{empheq}{alignat=7}\label{eq: WZ}
	F\wedge P[C_{(4)}]&=4F_{\rho\tau}f_2(\Theta)\sin^2\alpha\cos\alpha\sin\vartheta\,d^6\sigma\,.
\end{empheq}
The function $f_2(\Theta)$ is given in \eqref{eq: RR 4-form functions}.
\subsection{Supersymmetry}
The second-order equations derived from the DBI and WZ actions are difficult to solve. Instead of dealing with them directly we will require that the D5-brane configuration preserve the same $8$ supercharges as its F1 and D3-brane counterparts. This will lead to a set of first order equations which are easier to integrate. On general grounds we expect that any solution to the BPS equations is also a solution of the Euler-Lagrange equations.

A given D5-brane configuration will preserve some amount of supersymmetry if there exist target space Killing spinors satisfying
(see appendix \ref{app: conventions} for our spinor conventions)
\begin{empheq}{alignat=7}\label{eq: D5 BPS equation 1}
	\Gamma_{D5}\epsilon&=\epsilon\,,
\end{empheq}
where the $\kappa$-symmetry projector is \cite{Martucci:2005rb} (adapted to Euclidean signature)
\begin{empheq}{alignat=7}\label{eq: D5 projector 1}
	\Gamma_{D5}&=\frac{\epsilon^{abcdef}\sigma_2}{\sqrt{\det\left(g+F\right)}}\left(\frac{1}{6!}\Gamma_{abcdef}\sigma_3-\frac{1}{2\cdot4!}F_{ab}\Gamma_{cdef}+\frac{1}{8\cdot2!}F_{ab}F_{cd}\Gamma_{ef}\sigma_3-\frac{1}{48}F_{ab}F_{cd}F_{ef}\right)\,.
\end{empheq}
Here $\Gamma_a=\partial_ax^m\Gamma_m$ is the pullback of the $10$-dimensional Dirac matrices, $\sigma_i$ are Pauli matrices, and $\epsilon^{\rho\tau\alpha\psi\vartheta\varphi}=1$. In order to write this projector explicitly it is useful to introduce gamma matrices associated to the metric $h_{ab}$, namely,
\begin{empheq}{alignat=7}
	\gamma_{\rho}&=\Gamma_{\underline{1}}\,,
	&\qquad
	\gamma_{\tau}&=\sinh\rho\,\Gamma_{\underline{2}}\,,
	&\qquad
	\gamma_{\alpha}&=\sin\Theta\,\Gamma_{\underline{6}}\,,
	\cr
	\gamma_{\psi}&=\sin\Theta\cos\alpha\,\Gamma_{\underline{7}}\,,
	&\qquad
	\gamma_{\vartheta}&=\sin\Theta\sin\alpha\,\Gamma_{\underline{8}}\,,
	&\qquad
	\gamma_{\varphi}&=\sin\Theta\sin\alpha\sin\vartheta\,\Gamma_{\underline{9}}\,.
\end{empheq}
Then
\begin{empheq}{alignat=7}
	\Gamma_a&=\gamma_a+\partial_a\Theta\Gamma_{\underline{5}}\,,
\end{empheq}
and after some algebra we arrive at\footnote{The following identities are useful:
\begin{empheq}{alignat*=7}
	\frac{\epsilon^{abcdef}}{5!\sqrt{\det\left(h\right)}}\gamma_{bcdef}&=\gamma^a\gamma\,,
	&\quad
	\frac{\epsilon^{abcdef}}{4!\sqrt{\det\left(h\right)}}\gamma_{cdef}&=-\gamma^{ab}\gamma\,,
	&\quad
	\frac{\epsilon^{abcdef}}{3!\sqrt{\det\left(h\right)}}\gamma_{def}&=-\gamma^{abc}\gamma\,,
	&\quad
	\frac{\epsilon^{abcdef}}{2!\sqrt{\det\left(h\right)}}\gamma_{ef}&=\gamma^{abcd}\gamma\,.
\end{empheq}
}
\begin{empheq}{alignat=7}
	\Gamma_{D5}&=\frac{\gamma\sigma_2}{\sqrt{\mathcal{L}}}\left(\left(1+\gamma^a\partial_a\Theta\Gamma_{\underline{5}}\right)\sigma_3+\frac{1}{2}F_{ab}\left(\gamma^{ab}+\gamma^{abc}\partial_c\Theta\Gamma_{\underline{5}}\right)+\frac{1}{8}F_{ab}F_{cd}\gamma^{abcd}\sigma_3\right)\,,
\end{empheq}
with
\begin{empheq}{alignat=7}\label{eq: D5 projector 2}
	\gamma&\equiv\frac{\epsilon^{abcdef}}{6!\sqrt{\det\left(h\right)}}\gamma_{abcdef}\,.
\end{empheq}
The last term in \eqref{eq: D5 projector 1} vanishes since the gauge field effectively lives in 4 dimensions. In appendix \ref{app: explicit expressions} we write the expanded form of the projector $\Gamma_{D5}$.

Now, the dependence of the $AdS_5\times S^5$ Killing spinors \eqref{eq: Killing spinor} on the relevant coordinates is
\begin{empheq}{alignat=7}\label{eq: Killing spinor pullback}
	\epsilon&=e^{-\frac{i}{2}\rho\Gamma_*\Gamma_{\underline{1}}}e^{-\frac{i}{2}\Theta\Gamma_*\Gamma_{\underline{5}}}e^{\frac{1}{2}\alpha\Gamma_{\underline{56}}}e^{\frac{1}{2}\left(\tau\Gamma_{\underline{12}}+\psi\Gamma_{\underline{57}}\right)}M\epsilon_0\,,
    &\qquad
    M&=e^{\frac{1}{2}\eta\Gamma_{\underline{03}}}e^{\frac{1}{2}\xi\Gamma_{\underline{34}}}e^{\frac{1}{2}\vartheta\Gamma_{\underline{68}}}e^{\frac{1}{2}\varphi\Gamma_{\underline{89}}}\,,
\end{empheq}
where $\epsilon_0$ is a doublet of constant Weyl spinors. Borrowing from the supersymmetry analysis of the string solution (see appendix \ref{app: F1}) we impose the constraints
\begin{empheq}{alignat=7}\label{eq: D5 BPS constraints}
	\Gamma_{\underline{1257}}\epsilon_0&=\epsilon_0\,,
	&\qquad
	i\Gamma_{\underline{12}}e^{\theta_0\Gamma_{\underline{27}}}\sigma_3\epsilon_0&=\epsilon_0\,,
\end{empheq}
each of which reduces the number of preserved supercharges by half. Since the matrices $\Gamma_{\underline{1257}}$ and $i\Gamma_{\underline{12}}e^{\theta_0\Gamma_{\underline{27}}}\sigma_3$ commute with $M$ (and with each other), the spinors $\epsilon_0$ and $M\epsilon_0$ satisfy the same constraints. This preserves the $SO(3)\times SO(3)$ symmetry of the ansatz, as the dependence on the $S^2\subset AdS_5$ and $S^2\subset S^4$ coordinates carried by the matrix $M$ drops out from the projection \eqref{eq: D5 BPS equation 1}. Similarly, the first condition in \eqref{eq: D5 BPS constraints} implies that the Killing spinor \eqref{eq: Killing spinor pullback} depends on the difference $\Delta=\tau-\psi$, which is required by the $U(1)$ symmetry. 

Using the explicit form of the Killing spinors the BPS equation may be rewritten as
\begin{empheq}{alignat=7}
	U^{-1}\Gamma_{D5}UM\epsilon_0&=M\epsilon_0\,,
	&\qquad
	U&=e^{-\frac{i}{2}\rho\Gamma_*\Gamma_{\underline{1}}}e^{-\frac{i}{2}\Theta\Gamma_*\Gamma_{\underline{5}}}e^{\frac{1}{2}\alpha\Gamma_{\underline{56}}}e^{\frac{1}{2}\Delta\Gamma_{\underline{12}}}\,.
\end{empheq}
The matrix $U^{-1}\Gamma_{D5}U$ can now be expanded in the basis of totally antisymmetric products of Dirac matrices tensored with the $2\times2$ matrices $\sigma_i$ (we include $\sigma_0=\mathds{1}_{2\times2}$), that is,
\begin{empheq}{alignat=7}
	U^{-1}\Gamma_{D5}U&=c_i\,\mathds{1}_{32\times32}\,\sigma_i+c_{\left(\underline{m},i\right)}\,\Gamma^{\underline{m}}\,\sigma_i+c_{\left(\underline{mn},i\right)}\,\Gamma^{\underline{mn}}\,\sigma_i+c_{\left(\underline{mnp},i\right)}\,\Gamma^{\underline{mnp}}\,\sigma_i+\cdots\,.
\end{empheq}
In principle there are $2^{10}\times 4=4096$ terms in the expansion, but the constraints \eqref{eq: D5 BPS constraints} and the Weyl condition $\Gamma_{11}\epsilon_0=\epsilon_0$ reduce the number of independent matrices down to $512$. Since we do not want to impose any further constraints on $\epsilon_0$, all $512$ coefficients but $c_0$ must vanish. In turn, the coefficients can be computed by multiplying $U^{-1}\Gamma_{D5}U$ by the corresponding basis element and taking the trace. Using Maple and Mathematica we find that only 16 coefficients are non-zero, leading to the set of equations
\\~\\
\begin{subequations}
\begin{minipage}{0.5\textwidth}
\begin{empheq}{alignat=7}\label{eq: eq0}
    \mathds{1}_{32\times32}\,\sigma_0&
	\quad\longrightarrow\quad&
	\textrm{eq}_0&=\sqrt{\mathcal{L}}\,,
    \\
    \Gamma_{\underline{12}}\,\sigma_0&
	\quad\longrightarrow\quad&
	\textrm{eq}_1&=0\,,
    \\
	\Gamma_{\underline{16}}\,\sigma_0&
	\quad\longrightarrow\quad&
	\textrm{eq}_2&=0\,,
	\\
	\Gamma_{\underline{17}}\,\sigma_0&
	\quad\longrightarrow\quad&
	\textrm{eq}_3&=0\,,	
	\\
	\Gamma_{\underline{26}}\,\sigma_0&
	\quad\longrightarrow\quad&
	\textrm{eq}_4&=0\,,
    \\
	\Gamma_{\underline{27}}\,\sigma_0&
	\quad\longrightarrow\quad&
	\textrm{eq}_5&=0\,,
    \\
	\Gamma_{\underline{67}}\,\sigma_0&
	\quad\longrightarrow\quad&
	\textrm{eq}_6&=0\,,
    \\
    \Gamma_{\underline{0134}}\,\sigma_2&
	\quad\longrightarrow\quad&
	\textrm{eq}_7&=0\,,
\end{empheq}
\end{minipage}
\begin{minipage}{0.5\textwidth}
\begin{empheq}{alignat=7}
	\Gamma_{\underline{0234}}\,\sigma_2&
	\quad\longrightarrow\quad&
	\textrm{eq}_8&=0\,,
    \\
	\Gamma_{\underline{0346}}\,\sigma_2&
	\quad\longrightarrow\quad&
	\textrm{eq}_9&=0\,,
	\\
	\Gamma_{\underline{0347}}\,\sigma_2&
	\quad\longrightarrow\quad&
	\textrm{eq}_{10}&=0\,,
    \\
	\Gamma_{\underline{1267}}\,\sigma_0&
	\quad\longrightarrow\quad&
	\textrm{eq}_{11}&=0\,,
    \\
	\Gamma_{\underline{012346}}\,\sigma_2&
	\quad\longrightarrow\quad&
	\textrm{eq}_{12}&=0\,,
	\\
	\Gamma_{\underline{012347}}\,\sigma_2&
	\quad\longrightarrow\quad&
	\textrm{eq}_{13}&=0\,,
    \\
	\Gamma_{\underline{013467}}\,\sigma_2&
	\quad\longrightarrow\quad&
	\textrm{eq}_{14}&=0\,,
	\\
    \label{eq: eq15}
	\Gamma_{\underline{023467}}\,\sigma_2&
	\quad\longrightarrow\quad&
	\textrm{eq}_{15}&=0\,.
\end{empheq}
\end{minipage}
\end{subequations}
\\~\\
The expressions for $\left(\textrm{eq}_0,\textrm{eq}_1,\ldots,\textrm{eq}_{15}\right)$ are collected in appendix \ref{app: explicit expressions}. To arrive at these we have chosen to eliminate the matrices $\Gamma_{\underline{5}}$, $\sigma_1$ and $\sigma_3$ using the constraints \eqref{eq: D5 BPS constraints}. Similarly, the Weyl condition allowed us to replace $\Gamma_{\underline{89}}$ in favor of $\Gamma_{\underline{01234567}}\to\Gamma_{\underline{0346}}$.
\subsection{Solution}
The BPS conditions derived above form a set of 16 algebraic equations for the 6 variables
\begin{empheq}{alignat=7}
	\partial_{\rho}\theta\,,
	\qquad
	F_{\rho\tau}\,,
	\qquad
	F_{\rho\alpha}\,,
	\qquad
	F_{\rho\psi}\,,
	\qquad
	F_{\tau\alpha}\,,
	\qquad
    F_{\tau\psi}\,.
\end{empheq}
These equations are consistent with each other and, despite being quadratic, have a unique and remarkably simple solution. Indeed, using Maple we can eliminate the gauge field and solve for
\begin{empheq}{alignat=7}\label{eq: dt}
    \partial_{\rho}\theta&=-\frac{\cos\theta_0\sin\theta\cosh\rho-\sin\theta_0\cos\theta\sinh\rho}{\cos\theta_0\cos\theta\sinh\rho+\sin\theta_0\sin\theta\cosh\rho}\,.
\end{empheq}
To our surprise, this is the same equation that appears for the string configuration dual to the latitude Wilson loop in the fundamental representation of $SU(N)$ (cf. \eqref{eq: string BPS equation 5}-\eqref{eq: theta F1}). Demanding that $\theta\to0$ at the center of the $AdS_2$ disk,\footnote{As for the F1 and D3-brane, there is a second (unstable) solution given by
\begin{empheq}{alignat*=7}
	\sin\theta&=\frac{\sin\theta_0\sinh\rho}{\cosh\rho-\cos\theta_0}\,,
	&\qquad
    \cos\theta&=\frac{\cos\theta_0\cosh\rho-1}{\cosh\rho-\cos\theta_0}\,,
    &\qquad
    \theta_0&\leq\theta&&\leq\pi\,.
\end{empheq}
We do not explore this possibility here.
} 
as required by regularity of the induced geometry (more on this below), the solution to \eqref{eq: dt} is
\begin{empheq}{alignat=7}\label{eq: theta D5}
	\sin\theta&=\frac{\sin\theta_0\sinh\rho}{\cosh\rho+\cos\theta_0}\,,
	&\qquad
    \cos\theta&=\frac{\cos\theta_0\cosh\rho+1}{\cosh\rho+\cos\theta_0}\,,
    &\qquad
    0&\leq\theta&&\leq\theta_0\,.
\end{empheq}
The field strength then simplifies to
\begin{empheq}{alignat=7}
	\label{eq: F12}
	F_{\rho\tau}&=i\partial_{\rho}\left(\cos\Theta-\cos\theta_k\cosh\rho\right)+i\partial_{\tau}\left(\sin\Theta\cos\alpha\sin\Delta\right)\partial_{\rho}\theta\,,
	\\
	\label{eq: F16}
	F_{\rho\alpha}&=i\partial_{\alpha}\left(\sin\Theta\cos\alpha\sin\Delta\right)\partial_{\rho}\theta\,,
	\\
	\label{eq: F17}
	F_{\rho\psi}&=i\partial_{\psi}\left(\sin\Theta\cos\alpha\sin\Delta\right)\partial_{\rho}\theta\,,
	\\
	\label{eq: F26}
	F_{\tau\alpha}&=-i\partial_{\alpha}\left(\cos\Theta-\cos\theta_k\cosh\rho\right)\,,
	\\
	\label{eq: F27}
	F_{\tau\psi}&=-i\partial_{\psi}\left(\cos\Theta-\cos\theta_k\cosh\rho\right)\,.
\end{empheq}
Recall that the derivatives $\partial_a\Theta$ are give in \eqref{eq: R}-\eqref{eq: T}. Happily, these expressions satisfy the Bianchi identity $dF=0$ and can be derived from the potential
\begin{empheq}{alignat=7}\label{eq: gauge field potential}
	A&=-i\sin\Theta\cos\alpha\sin\Delta\,d\theta+i\left(\cos\Theta-\cos\theta_k\cosh\rho\right)d\tau\,.
\end{empheq}
This configuration correctly reproduces the $\frac{1}{2}$-BPS case \eqref{eq: 1/2-BPS solution} in the $\theta_0\to0$ limit. We have also checked that it satisfies the second order Euler-Lagrange equations.

To study the regularity of the solution we invoke the implicit function theorem. Define
\begin{empheq}{alignat=7}\label{eq: beta}
    \beta&\equiv\cos\theta\cos\Theta+\sin\theta\sin\Theta\cos\alpha\cos\Delta\,.
\end{empheq}
When evaluated on the surface \eqref{eq: beta=cos theta_k} we find that
\begin{empheq}{alignat=7}
	\partial^m\beta\partial_m\beta&=\left[\left(\sin\theta\cos\Theta-\cos\theta\sin\Theta\cos\alpha\cos\Delta\right)^2+\sin^2\Theta\cos^2\alpha\sin^2\Delta\right]\left(\partial_{\rho}\theta\right)^2+\sin^2\theta_k\,.
\end{empheq}
This is manifestly positive, so the geometry is smooth. Notice, however, that the static gauge coordinates $\sigma^a=(\rho,\tau,\alpha,\psi,\vartheta,\varphi)$ do not cover the entire manifold, as they fail to include points where
\begin{empheq}{alignat=7}\label{eq: singular coordinates}
	\partial_{\Theta}\beta&=0
    &\qquad\Rightarrow\qquad
    \cos\theta\sin\Theta-\sin\theta\cos\Theta\cos\alpha\cos\Delta&=0
	&\qquad\Rightarrow\qquad
	\cos\theta&=\cos\theta_k\cos\Theta\,.
\end{empheq}
If $\theta_0$ lies inside the range
\begin{empheq}{alignat=7}\label{eq: regularity range}
	\textrm{min}\left(\theta_k,\pi-\theta_k\right)&<\theta_0&&<\textrm{max}\left(\theta_k,\pi-\theta_k\right)\,,
\end{empheq}
then the coordinates can become singular and one must choose a different parametrization for the surface $\beta=\cos\theta_k$. Still, the induced metric is regular everywhere. Regarding the gauge field, the 1-forms $d\rho$, $d\tau$ and $\rho\,d\tau$ are ill-defined at $\rho=0$, so we need to study the behavior of the solution near the center of the $AdS_2$ disk. We can solve equation \eqref{eq: beta=cos theta_k} perturbatively in $\rho$ to find\footnote{For small enough $\rho$ condition \eqref{eq: singular coordinates} is never satisfied, so the coordinates are regular.}
\begin{empheq}{alignat=7}
	\Theta&=\theta_k+\frac{\sin\theta_0\cos\alpha\cos\Delta}{1+\cos\theta_0}\rho-\frac{\cot\theta_k\sin^2\theta_0\left(1-\cos^2\alpha\cos^2\Delta\right)}{2\left(1+\cos\theta_0\right)^2}\rho^2+\mathcal{O}\left(\rho^3\right)\,.
\end{empheq}
The induced metric \eqref{eq: induced metric} and the gauge field \eqref{eq: gauge field potential} then read
\begin{empheq}{alignat=7}
	ds^2&=d\rho^2+\rho^2d\tau^2+\frac{\sin^2\theta_0\cos^2\alpha}{\left(1+\cos\theta_0\right)^2}\left(\cos\Delta\,d\rho-\rho\sin\Delta\,d\tau\right)^2+\cdots\,,
	\\
	A&=-\frac{i\sin\theta_k\sin\theta_0\cos\alpha}{1+\cos\theta_0}\left(\sin\Delta\,d\rho+\rho\cos\Delta\,d\tau\right)+\cdots\,.
\end{empheq}
The dots represent terms that are regular as $\rho\to0$ (e.g. $\rho\,d\rho$, $\rho^2d\tau$). Switching to cartesian coordinates $(x,y)=(\rho\cos\tau,\rho\sin\tau)$ this becomes
\begin{empheq}{alignat=7}
	ds^2&=dx^2+dy^2+\frac{\sin^2\theta_0\cos^2\alpha}{\left(1+\cos\theta_0\right)^2}\left(\cos\psi\,dx+\sin\psi\,dy\right)^2+\cdots\,,
	\\
	A&=-\frac{i\sin\theta_k\sin\theta_0\cos\alpha}{1+\cos\theta_0}\left(-\sin\psi\,dx+\cos\psi\,dy\right)+\cdots\,.
\end{empheq}
Both fields are manifestly regular at $\rho=0$ in these coordinates. Of course, $A$ can become singular after a gauge transformation, but the curvature $F$ will remain smooth.
\section{String charge and on-shell action}\label{sec: on-shell}
Having found the solution to the BPS equations we now compute the on-shell action and the string charge carried by the D5-brane. To this purpose we first point out that the DBI Lagrangian \eqref{eq: DBI} simplifies to
\begin{empheq}{alignat=7}\label{eq: on-shell L}
	\mathcal{L}&=\frac{\sin^2\Theta\left(\sin^2\theta_k-\sin^2\Theta\sin^2\alpha\,(\partial_{\rho}\theta)^2\right)^2}{\left(\cos\theta-\cos\theta_k\cos\Theta\right)^2}\,.
\end{empheq}
As usual in supersymmetric setups, this is a perfect square. Other useful simplifications are
\begin{empheq}{alignat=7}\label{eq: on-shell Pa}
	\frac{\partial\mathcal{L}}{\partial F_{\rho\tau}}&=-\frac{2i}{\sinh\rho}\frac{\cos\theta_k\sin^2\Theta\left(\sin^2\theta_k-\sin^2\Theta\sin^2\alpha\,(\partial_{\rho}\theta)^2\right)}{\left(\cos\theta-\cos\theta_k\cos\Theta\right)^2}\,,
\end{empheq}
and
\begin{empheq}{alignat=7}\label{eq: on-shell F12}
    F_{\rho\tau}&=-i\sinh\rho\left(\cos\theta_k-\frac{\cos\Theta\sin^2\Theta\sin^2\alpha\,(\partial_{\rho}\theta)^2}{\cos\theta-\cos\theta_k\cos\Theta}\right)\,.
\end{empheq}
We anticipate, however, that we have been unable to perform the next calculations exactly in $\theta_0$, so we proceed perturbatively using the series solution
\begin{empheq}{alignat=7}\label{eq: perturbative solution}
	\Theta&=\theta_k+\frac{\sinh\rho\cos\alpha\cos\Delta}{\cosh\rho+1}\theta_0-\frac{\cot\theta_k\sinh^2\rho\left(1-\cos^2\alpha\cos^2\Delta\right)}{2\left(\cosh\rho+1\right)^2}\theta_0^2+\cdots\,.
\end{empheq}
Higher order terms are easily obtained from \eqref{eq: beta=cos theta_k} using Maple or Mathematica. In what follows we only present the details to order $\theta_0^2$, but we have actually done the calculations up to $\mathcal{O}(\theta_0^{10})$.

The string charge dissolved on the D5-brane is equal to the electric charge that sources the gauge field $F_{ab}$. It can be computed using Gauss's Law, which in our coordinates takes the form
\begin{empheq}{alignat=7}\label{eq: string charge definition}
	k&=2\pi\alpha'i\int_{\rho\,\to\infty}d\alpha d\psi d\vartheta d\varphi\,\Pi_A\,,
	&\qquad
	\Pi_A&=\frac{\delta S_{D5}}{\delta F_{\rho\tau}}\,.
\end{empheq}
Here we have reinstated the factor of $2\pi\alpha'$ that was previously absorbed in $F_{ab}$. The $i$ is due to the Euclidean continuation. Using \eqref{eq: on-shell L} and \eqref{eq: on-shell Pa} the string charge simplifies to
\begin{empheq}{alignat=7}\label{eq: string charge integral}
	k&=\frac{N}{\pi^2}\int_{\rho\,\to\infty} d\alpha d\Delta\,\sin^2\alpha\cos\alpha\left(\frac{\cos\theta_k\sin^5\Theta}{\cos\theta_0-\cos\theta_k\cos\Theta}+\frac{3\Theta}{2}-\sin(2\Theta)+\frac{1}{8}\sin(4\Theta)\right)\,,
\end{empheq}
and plugging in the perturbative solution \eqref{eq: perturbative solution} this becomes
\begin{empheq}{alignat=7}
	k&=\frac{N}{\pi^2}\int_{\rho\,\to\infty} d\alpha d\Delta\,\sin^2\alpha\cos\alpha\left[\frac{3}{2}\left(\theta_k-\cos\theta_k\sin\theta_k\right)+4\sin^2\theta_k\cos\alpha\cos\Delta\,\theta_0\right.
	\cr
	&\left.+\frac{3}{2}\cos\theta_k\sin\theta_k\left(5\cos^2\alpha\cos^2\Delta-1\right)\theta_0^2+\cdots\right]\,.
\end{empheq}
The second term clearly vanishes after integration. What is not so obvious is that the third term also vanishes. In fact, we have checked using Maple that the integral \eqref{eq: string charge integral} gives
\begin{empheq}{alignat=7}\label{eq: string charge 1/4 BPS}
	k&=\frac{N}{\pi}\left(\theta_k-\cos\theta_k\sin\theta_k\right)+\mathcal{O}\left(\theta_0^{10}\right)\,,
\end{empheq}
leading us to conjecture that it is independent of $\theta_0$. Thus, the relation between the integration constant $\theta_k$ and the string charge $k$ is the same as in the $\frac{1}{2}$-BPS case.

The calculation of the on-shell action \eqref{eq: D5 action} is more subtle since it is divergent. Indeed, from \eqref{eq: on-shell L} and \eqref{eq: on-shell F12} one obtains
\begin{empheq}{alignat=7}
    S_{D5}&=\frac{N\sqrt{\lambda}}{\pi^2}\int d\rho d\alpha d\Delta\,\sinh\rho\sin^2\alpha\cos\alpha\Bigg[\frac{\sin^5\Theta\left(\sin^2\theta_k-\sin^2\Theta\sin^2\alpha\,(\partial_{\rho}\theta)^2\right)}{\cos\theta-\cos\theta_k\cos\Theta}
    \cr
    &-\left(\cos\theta_k-\frac{\cos\Theta\sin^2\Theta\sin^2\alpha\,(\partial_{\rho}\theta)^2}{\cos\theta-\cos\theta_k\cos\Theta}\right)\left(\frac{3\Theta}{2}-\sin(2\Theta)+\frac{1}{8}\sin(4\Theta)\right)\Bigg]
    \cr
    &=\frac{N\sqrt{\lambda}}{\pi}\left(\cos\theta_k\left(\theta_k-\cos\theta_k\sin\theta_k\right)-\frac{2}{3}\sin^3\theta_k\right)\left(-\frac{1}{2}e^{\rho_c}+1+\frac{3}{10}\theta_0^2+\cdots\right)\,,
\end{empheq}
where $0<\rho<\rho_c\to\infty$ is a large cutoff. The main points to highlight are that the result factorizes as shown above, the divergent piece is independent of $\theta_0$, and odd powers in the expansion vanish after integration, at least to order $\mathcal{O}(\theta_0^{10})$. Now, the standard prescription to renormalize the action is to perform Legendre transforms on some of the worldvolume fields \cite{Yamaguchi:2006tq,Drukker:2006zk,Drukker:1999zq,Drukker:2005kx}. Concretely, we first add the boundary term
\begin{empheq}{alignat=7}\label{eq: Legendre A}
	S^{\textrm{bdry}}_A&=-\int_{\rho\,=\rho_c}d\tau d\alpha d\psi d\vartheta d\varphi\,\Pi_AA_{\tau}\,,
\end{empheq}
which, from the variational point of view, fixes the total electric charge on the brane as opposed to the value of the potential at the boundary. This is natural in our context since the AdS/CFT dictionary maps $k$ to the rank of the representation $\mathcal{A}_k$ of $SU(N)$. Using the perturbative solution we find
\begin{empheq}{alignat=7}
	S^{\textrm{bdry}}_A&=\frac{N\sqrt{\lambda}}{\pi^2}\int_{\rho\,=\rho_c}d\alpha d\Delta\,\sin^2\alpha\cos\alpha\left(\frac{\cos\theta_k\sin^5\Theta}{\cos\theta_0-\cos\theta_k\cos\Theta}+\frac{3\Theta}{2}-\sin(2\Theta)+\frac{1}{8}\sin(4\Theta)\right)
    \cr
    &\times\left(\cos\theta_k\cosh\rho-\cos\Theta\right)
    \cr
	&=\frac{N\sqrt{\lambda}}{\pi}\left[\cos\theta_k\left(\theta_k-\cos\theta_k\sin\theta_k\right)\left(\frac{1}{2}e^{\rho_c}-1-\frac{3}{10}\theta_0^2\right)+\frac{8}{15}\sin^3\theta_k\theta_0^2+\cdots\right]\,.
\end{empheq}
Again, odd powers vanish and the divergence is $\theta_0$-independent. The second part of the renormalization prescription is to perform a Legendre transform on the $z$ coordinate of the D5-brane embedding in the Poincar\'e patch of $AdS_5$, which requires adding the term
\begin{empheq}{alignat=7}
	S^{\textrm{bdry}}_z&=-\int_{\rho\,=\rho_c}d\tau d\alpha d\psi d\vartheta d\varphi\,\Pi_zz\,,
	&\qquad
	\Pi_z&=\frac{\delta S_{D5}}{\delta z'}\,.
\end{empheq}
This was justified in \cite{Drukker:1999zq,Drukker:2005kx} in terms of the Neumann boundary conditions satisfied by open strings in the directions parallel to the D3-branes that backreact to the $AdS_5\times S^5$ geometry. Given that $z\approx e^{-2\rho}$ close to the boundary, the above is equivalent to
\begin{empheq}{alignat=7}
	S^{\textrm{bdry}}_z&=-\int_{\rho\,=\rho_c}d\tau d\alpha d\psi d\vartheta d\varphi\,\Pi_{\rho}\,,
	&\qquad
	\Pi_{\rho}&=\frac{\delta S_{D5}}{\delta\rho'}\,.
\end{empheq}
To compute the momentum conjugate to $\rho$ we must undo the static gauge-fixing and introduce a new worldvolume coordinate $\sigma$ such that $\rho=\rho(\sigma)$. This amounts to replacing the first component of the metric $h_{ab}$ in \eqref{eq: h metric} by $h_{\rho\rho}=1\to h_{\sigma\sigma}=\rho'^2$ and defining the Lagrangian \eqref{eq: DBI} using this new metric (see appendix \ref{app: explicit expressions}). After computing $\Pi_{\rho}$ we can set $\rho'=1$. With the help of Maple we get
\begin{empheq}{alignat=7}
	S^{\textrm{bdry}}_z&=-\frac{N\sqrt{\lambda}}{\pi}\int_{\rho\,=\rho_c}d\alpha d\Delta\,\frac{\sinh\rho\sin^2\alpha\cos\alpha\sin^5\Theta}{\cos\theta-\cos\theta_k\cos\Theta}
    \cr
    &=-\frac{N\sqrt{\lambda}}{3\pi}\sin^3\theta_k\,e^{\rho_c}+\mathcal{O}\left(\theta_0^{10}\right)\,.
\end{empheq}
As we can see, this term does not contribute with a finite piece and is independent of $\theta_0$, at least to the order shown above. Putting all the ingredients together, we find that the renormalized action for the $\frac{1}{4}$-BPS D5-brane is
\begin{empheq}{alignat=7}
	S_{D5}+S^{\textrm{bdry}}_z+S^{\textrm{bdry}}_A&=-\frac{2N\sqrt{\lambda}}{3\pi}\sin^3\theta_k\left(1-\frac{1}{2}\theta_0^2+\cdots\right)\,.
\end{empheq}
This is finite. Moreover, we have checked to order $\mathcal{O}\left(\theta_0^{10}\right)$ that
\begin{empheq}{alignat=7}\label{eq: on-shell action}
    S_{D5}+S^{\textrm{bdry}}_z+S^{\textrm{bdry}}_A&=-\frac{2N\sqrt{\lambda}}{3\pi}\sin^3\theta_k\cos\theta_0\,,
\end{empheq}
which coincides with the gauge theory prediction \eqref{eq: vev A} according to the holographic dictionary.
\section{Conclusions}\label{sec: conclusions}
In this paper we have found the D5-brane configuration dual to the $\frac{1}{4}$-BPS latitude Wilson loops in the $k$-antisymmetric representation of $SU(N)$, thus completing a missing entry in the AdS/CFT dictionary. Our solution is exact in the latitude parameter $\theta_0$ and correctly reproduces the $\frac{1}{2}$-BPS limit. Unfortunately, we only managed to compute the string charge and on-shell action perturbatively. We found full agreement with the gauge theory result to order $\mathcal{O}\left(\theta_0^{10}\right)$.

\begin{figure}[b]
     \centering
     \begin{subfigure}[b]{0.475\textwidth}
         \centering
         \includegraphics[width=\textwidth]{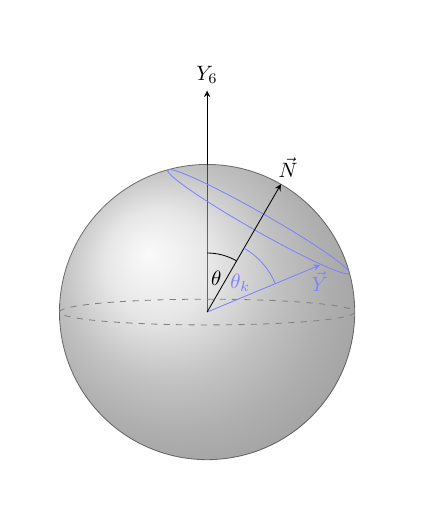}
         \caption{The $\frac{1}{4}$-BPS D5-brane satisfies $\vec{Y}\cdot\vec{N}=\cos\theta_k$.}
     \end{subfigure}
     \hfill
     \begin{subfigure}[b]{0.475\textwidth}
         \centering
         \includegraphics[width=\textwidth]{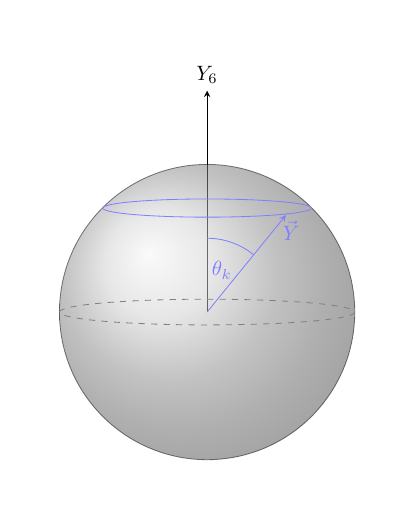}
         \caption{The $\theta_0\to0$ limit  reproduces the $\frac{1}{2}$-BPS case.}
     \end{subfigure}
        \caption{Viewed in embedding coordinates $\vec{Y}\in\mathds{R}^6$, the D5-brane wraps an $S^4\subset S^5$ corresponding to a fixed polar angle $\theta_k$ measured with respect to the axis $\vec{N}$, which itself depends on the $AdS_2$ coordinates $(\rho,\tau)$. The axis starts at the north pole of the $5$-sphere for $\rho=0$ and reaches $\theta=\theta_0$ at the boundary $\rho\to\infty$; as $\tau$ varies it rotates around the north pole. The angle $\theta_k$ is determined by the value of the string charge $k$ carried by the D5-brane.}
        \label{fig: D5}
\end{figure}

Referring to the coordinate system \eqref{eq: AdS5 metric}-\eqref{eq: S5 coordinate range}, the D5-brane spans the $AdS_2\subset AdS_5$ disk located at $u=0$ while wrapping an $S^4\subset S^5$ corresponding to a constant polar angle $0\leq\theta_k\leq\pi$ measured with respect to the axis $\vec{N}\in\mathds{R}^6$ given in \eqref{eq: axis N} and \eqref{eq: theta D5}. This is represented in figure \ref{fig: D5}. It is important to emphasize that because the axis depends on the worldvolume coordinates $(\rho,\tau)$, the D5-brane wraps a different $4$-sphere at each point in $AdS_2$. On the other hand, the value of the angle $\theta_k$ is determined by the string charge $k$ dissolved on the brane according to formula \eqref{eq: string charge 1/4 BPS}. Even though we computed this relation perturbatively, we conjecture it to be independent of the latitude parameter $\theta_0$ and therefore identical to the $\frac{1}{2}$-BPS version \eqref{eq: string charge 1/2 BPS}. In turn, the electric charge $k$ sources the worldvolume gauge field \eqref{eq: F12}-\eqref{eq: F27}, derivable from the potential \eqref{eq: gauge field potential}. Notice that, unlike the D3-brane solution, which only carries an electric field $F_{\rho\tau}$, the D5-brane also has magnetic components.

As expected, the calculation of the D5-brane on-shell action \eqref{eq: on-shell action} required the regularization of divergences and their corresponding renormalization. The standard prescription of performing Legendre transforms on some of the worldvolume fields rendered the correct the result, as it did for the F1 and D3-brane configurations in \cite{Yamaguchi:2006tq,Drukker:2006zk,Drukker:2005kx}. A slight technical deviation from previous works, however, is that we implemented the transform of the electric component of the gauge field using the boundary term \eqref{eq: Legendre A}, as opposed to a bulk integral involving $F_{\rho\tau}$. This emphasizes the role that regularity plays in the evaluation of the on-shell action. Indeed, our method only works in a gauge in which $A_{\tau}$ is smooth at $\rho=0$. Otherwise, one must incise a disk of radius $\epsilon\ll1$ from the center of $AdS_2$ and include additional boundary terms at $\rho=\epsilon$. Of course, this is not an issue when working with $F_{\rho\tau}$ since it is always regular.

An interesting difference between the D5-brane solution and its F1 and D3-brane counterparts lies in the interpretation of the vector $\vec{N}$, which is an extension into the bulk of the scalar couplings \eqref{eq: latitude couplings} that define the latitude Wilson loop. For the F1 and D3-brane configurations one identifies $\vec{N}=\vec{Y}$, where $\vec{Y}$ are embedding coordinates for $S^5\subset\mathds{R}^6$. Instead, for the D5-brane, $\vec{N}$ is interpreted as an axis such that $\vec{N}\cdot\vec{Y}=\cos\theta_k$. Regardless of the interpretation, the $SO(3)$ symmetry preserved by the three solutions corresponds to the subgroup of $SO(6)$ that leaves $\vec{N}$ invariant for all values of $\rho$ and $\tau$. In the case of the D5-brane, this is realized as isometries that act on the $S^2\subset S^4$ coordinates $(\vartheta,\varphi)$, not touching $(\alpha,\psi)$. The F1 and D3-brane both sit at $\alpha=0$, so the $2$-sphere shrinks to zero size. Another difference worth pointing out is the way in which the $U(1)$ symmetry manifests itself. Again referring to \eqref{eq: AdS5 metric}-\eqref{eq: S5 coordinate range}, the angular dependence drops out from the F1 and D3-brane solutions because they have $\tau=\psi$, whereas the D5-brane embedding does depend on these angles but only through the difference $\Delta=\tau-\psi$. In particular, this means that the preserved Killing spinors \eqref{eq: Killing spinor pullback}-\eqref{eq: D5 BPS constraints} carry this dependence.

Remarkably, the function $\theta(\rho)$ that determines the orientation of $\vec{N}$ is the same in the F1 and D5-brane cases. This is reminiscent of \cite{Hartnoll:2006ib}, where it was shown that any string solution that sits at a fixed point in $S^5$ can be extended to a D5-brane configuration with the same embedding in $AdS_5$ while wrapping a fixed $S^4\subset S^5$. Our results suggests that it might be possible to generalize this to string worldsheets that have a non-trivial profile in $S^5$. Perhaps, this can be interpreted in terms of the Myers effect \cite{Myers:1999ps}, whereby the worldsheet of $k$ coincident strings expands into additional directions.

The spectrum of fluctuations of the $\frac{1}{4}$-BPS string dual to the latitude Wilson loop was computed in \cite{Forini:2015mca,Forini:2015bgo} and later fit into supermultiplets of $SU(2|2)$ \cite{Faraggi:2016ekd}. The same was done for the $\frac{1}{2}$-BPS configurations in \cite{Faraggi:2011bb,Faraggi:2011ge}, where they organized the excitations in terms of representations of $OSp(4^*|4)$. It would be interesting to repeat this exercise for the $\frac{1}{4}$-BPS D3- and D5-brane solutions, perhaps allowing for the computation of $1/N$ corrections to the expectation value of latitude Wilson loops as in \cite{Buchbinder:2014nia}. This would also open up another setup where to study correlation functions of insertions in 1-dimensional defects and AdS$_{2}$/dCFT$_{1}$ in higher representations, along the lines of \cite{Giombi:2020amn}. Another follow up work is to apply the technology of calibration forms \cite{Dymarsky:2006ve,Mezei:2018url,Drukker:2021vyx,Drukker:2022kuz} to compute the on-shell action exactly in $\theta_0$. Finally, it remains to elucidate the role of the unstable D5-brane solution. All these are interesting avenues to pursue in the future.
\section*{Acknowledgements}
We would like to thank Max Ba\~nados, Diego Correa, Guille Silva, and Nadav Drukker for useful comments and discussions. The work of AF and CM is supported by ANID/ACT210100 Anillo Grant “Holography and its applications to High Energy Physics, Quantum Gravity and Condensed Matter Systems.” AF would like to acknowledge support from the ICTP through the Associates Programme (2022-2027).
\appendix
\section{Conventions}\label{app: conventions}
We work in Euclidean signature. Target space indices are labeled by $m,n,\ldots$. Worldvolume indices are $a,b,\ldots$. Tangent space indices are underlined. The background metric is $G_{mn}$. Our spinor conventions follow \cite{Martucci:2005rb}. In particular, type IIB fermions are grouped into a doublet of Weyl spinors of positive chirality, namely,\footnote{In Lorentzian signature type IIB spinors are Majorana-Weyl. It then makes sense to define the singlet $\epsilon=\epsilon_1-i\epsilon_2$ and its charge conjugate $\epsilon^c=\epsilon_1+i\epsilon_2$. However, since the Majorana condition is lost in Euclidean signature, we prefer to maintain the doublet notation.}
\begin{empheq}{alignat=7}
	\epsilon&=\left(
	\begin{array}{c}
		\epsilon_1
		\\
		\epsilon_2
	\end{array}
	\right)\,,
	&\qquad
	\Gamma_{11}\epsilon_1&=\epsilon_1\,,
	&\qquad
	\Gamma_{11}\epsilon_2&=\epsilon_2\,,
	&\qquad
	\Gamma_{11}&\equiv-i\Gamma_{\underline{0123456789}}\,.
\end{empheq}
The $AdS_5\times S^5$ Killing spinors satisfy
\begin{empheq}{alignat=7}
	\left(\nabla_m+\frac{i}{2L}\Gamma_*\Gamma_m\right)\epsilon&=0\,,
	&\qquad
	\Gamma_*&\equiv-i\Gamma_{\underline{01234}}\sigma_2\,.
\end{empheq}
In the coordinate system \eqref{eq: AdS5 metric}-\eqref{eq: S5 metric} the solution reads (with the obvious vielbein and spin connection)
\begin{empheq}{alignat=7}\label{eq: Killing spinor}
	\epsilon&=e^{-\frac{i}{2}u\Gamma_*\Gamma_{\underline{0}}}e^{-\frac{i}{2}\rho\Gamma_*\Gamma_{\underline{1}}}e^{\frac{1}{2}\tau\Gamma_{\underline{12}}}e^{\frac{1}{2}\eta\Gamma_{\underline{03}}}e^{\frac{1}{2}\xi\Gamma_{\underline{34}}}e^{-\frac{i}{2}\Theta\Gamma_*\Gamma_{\underline{5}}}e^{\frac{1}{2}\alpha\Gamma_{\underline{56}}}e^{\frac{1}{2}\psi\Gamma_{\underline{57}}}e^{\frac{1}{2}\vartheta\Gamma_{\underline{68}}}e^{\frac{1}{2}\varphi\Gamma_{\underline{89}}}\epsilon_0\,,
\end{empheq}
where $\epsilon_0$ is a doublet of constant Weyl spinors.
\section{String solution}\label{app: F1}
In this appendix we derive and solve the BPS equations for the string configuration dual to the latitude Wilson loop in the fundamental representation of $SU(N)$. We work in a static gauge in which the worldsheet coordinates are $\sigma^a=\left(\rho,\tau\right)$. The simplest ansatz that respects the $U(1)\times SO(3)\times SO(3)$ symmetry is
\begin{empheq}{alignat=7}\label{eq: string ansatz}
	u&=0\,,
	&\qquad
	\Theta&=\theta(\rho)\,,
	&\qquad
	\alpha&=0\,,
	&\qquad
	\psi&=\tau\,,
\end{empheq}
with $\theta(\rho\to\infty)=\theta_0$. For simplicity we also set $\eta=\xi=\vartheta=\varphi=0$, although the final result does not depend on this choice. The condition for supersymmetry reads
\begin{empheq}{alignat=7}
	\Gamma_{F1}\epsilon&=\epsilon\,,
\end{empheq}
where the $\kappa$-symmetry projector for a fundamental string is \cite{Drukker:2000ep}
\begin{empheq}{alignat=7}\label{eq: string kappa-symmetry projector 1}
	\Gamma_{F1}&\equiv\frac{i\epsilon^{ab}\Gamma_{ab}\sigma_3}{2!\sqrt{\det\left(g\right)}}\,.
\end{empheq}
In this case we find
\begin{empheq}{alignat=7}\label{eq: string kappa-symmetry projector 2}
	\Gamma_{F1}&=\frac{i\left(\sinh\rho\left(\Gamma_{\underline{12}}-\partial_{\rho}\theta\,\Gamma_{\underline{25}}\right)+\sin\theta\left(\Gamma_{\underline{17}}+\partial_{\rho}\theta\,\Gamma_{\underline{57}}\right)\right)\sigma_3}{\sqrt{\left(1+(\partial_{\rho}\theta)^2\right)\left(\sinh^2\rho+\sin^2\theta\right)}}\,.
\end{empheq}
Notice that \eqref{eq: string kappa-symmetry projector 2} does not depend on $\tau$. Also, it does not commute with $\Gamma_{\underline{12}}$ and $\Gamma_{\underline{57}}$. This forces us to remove the $\tau$-dependence from the Killing spinor \eqref{eq: Killing spinor} by imposing
\begin{empheq}{alignat=7}\label{eq: string BPS condition 1}
	\left(\Gamma_{\underline{12}}+\Gamma_{\underline{57}}\right)\epsilon_0&=0
	&\qquad\Leftrightarrow\qquad
	\Gamma_{\underline{1257}}\epsilon_0&=\epsilon_0\,.
\end{empheq}
The supersymmetry condition then becomes
\begin{empheq}{alignat=7}\label{eq: string BPS equation 1}
	U^{-1}\Gamma_{F1}U\epsilon_0&=\epsilon_0\,,
	&\qquad
	U&=e^{-\frac{i}{2}\rho\Gamma_*\Gamma_{\underline{1}}}e^{-\frac{i}{2}\Theta\Gamma_*\Gamma_{\underline{5}}}\,.
\end{empheq}
Some algebra shows that
\begin{empheq}{alignat=7}
	U^{-1}\Gamma_{F1}U&=\frac{i\left(\sinh\rho\cos\theta\left(\Gamma_{\underline{12}}-\partial_{\rho}\theta\,\Gamma_{\underline{25}}\right)+\cosh\rho\sin\theta\left(\Gamma_{\underline{17}}+\partial_{\rho}\theta\,\Gamma_{\underline{57}}\right)\right)\sigma_3}{\sqrt{\left(1+(\partial_{\rho}\theta)^2\right)\left(\sinh^2\rho+\sin^2\theta\right)}}
    \cr
&+\frac{i\sinh\rho\sin\theta\Gamma_*\left(\Gamma_{\underline{125}}+\partial_{\rho}\theta\,\Gamma_{\underline{2}}+\Gamma_{\underline{7}}+\partial_{\rho}\theta\,\Gamma_{\underline{157}}\right)\sigma_3}{\sqrt{\left(1+(\partial_{\rho}\theta)^2\right)\left(\sinh^2\rho+\sin^2\theta\right)}}\,.
\end{empheq}
It is easy to see that the second line vanishes after imposing the constraint \eqref{eq: string BPS condition 1}. The projection then simplifies to
\begin{empheq}{alignat=7}\label{eq: string BPS equation 2}
    \frac{i\partial_{\rho}\left(\cosh\rho\cos\theta\,\Gamma_{\underline{12}}+\sinh\rho\sin\theta\,\Gamma_{\underline{17}}\right)\sigma_3}{\sqrt{\left(1+(\partial_{\rho}\theta)^2\right)\left(\sinh^2\rho+\sin^2\theta\right)}}\epsilon_0&=\epsilon_0\,.
\end{empheq}
Taking $\rho\to\infty$ we get a second condition on $\epsilon_0$, namely,
\begin{empheq}{alignat=7}\label{eq: string BPS condition 2}
	i\left(\cos\theta_0\Gamma_{\underline{12}}+\sin\theta_0\Gamma_{\underline{17}}\right)\sigma_3\epsilon_0&=\epsilon_0
	&\qquad\Leftrightarrow\qquad
	i\Gamma_{\underline{12}}e^{\theta_0\Gamma_{\underline{27}}}\sigma_3\epsilon_0&=\epsilon_0\,,
\end{empheq}
where we have assumed that $\partial_{\rho}\theta$ vanishes at the boundary. Importantly, $[i\Gamma_{\underline{12}}e^{\theta_0\Gamma_{\underline{27}}}\sigma_3,\Gamma_{\underline{1257}}]=0$, so the two constraints are compatible with each other, reducing the preserved supercharges from $32$ down to $8$. Inserting this back into \eqref{eq: string BPS equation 2} we get
\begin{empheq}{alignat=7}\label{eq: string BPS equation 3}
	\frac{\partial_{\rho}\left(\cosh\rho\cos\theta-\sinh\rho\sin\theta\,\Gamma_{\underline{27}}\right)e^{\theta_0\Gamma_{\underline{27}}}}{\sqrt{\left(1+(\partial_{\rho}\theta)^2\right)\left(\sinh^2\rho+\sin^2\theta\right)}}\epsilon_0&=\epsilon_0\,.
\end{empheq}
Expanding the exponential we obtain a term proportional to the identity matrix and another term proportional to $\Gamma_{\underline{27}}$. They lead to the pair of equations
\begin{empheq}{alignat=7}\label{eq: string BPS equation 4}
	\frac{\partial_{\rho}\left(\cos\theta_0\cosh\rho\cos\theta+\sin\theta_0\sinh\rho\sin\theta\right)}{\sqrt{\left(1+(\partial_{\rho}\theta)^2\right)\left(\sinh^2\rho+\sin^2\theta\right)}}&=1\,,
	\\ \label{eq: string BPS equation 5}
	\partial_{\rho}\left(\sin\theta_0\cosh\rho\cos\theta-\cos\theta_0\sinh\rho\sin\theta\right)&=0\,.
\end{empheq}
These are consistent with each other and imply
\begin{empheq}{alignat=7}
	\cosh\rho\cos\theta\sin\theta_0-\sinh\rho\sin\theta\cos\theta_0&=\pm\sin\theta_0\,.
\end{empheq}
The integration constant on the right hand side is fixed by looking at $\rho\to0$ and demanding that $\sin\theta\to0$, as required by regularity of the induced geometry. One can in fact solve for $\theta(\rho)$ to find
\begin{empheq}{alignat=7}\label{eq: theta F1}
	\sin\theta&=\frac{\sin\theta_0\sinh\rho}{\cosh\rho\pm\cos\theta_0}\,,
	&\qquad
	\cos\theta&=\frac{\cos\theta_0\cosh\rho\pm1}{\cosh\rho\pm\cos\theta_0}\,.
\end{empheq}
For the upper/lower sign the string wraps the northern/southern hemisphere of the $S^2\subset S^5$. Assuming that $0\leq\theta_0\leq\frac{\pi}{2}$, the stable solution corresponds to the $+$ sign.
\section{Explicit expressions}\label{app: explicit expressions}
Here we collect some explicit expressions omitted in the body of the paper. First, using a worldvolume coordinate $\sigma$ such that $\rho=\rho(\sigma)$, the DBI Lagrangian \eqref{eq: DBI} reads
\begin{empheq}{alignat=7}\label{eq: DBI 2}
	\mathcal{L}&=1+\frac{\left(\partial_{\sigma}\Theta\right)^2}{\rho'^2}+\frac{\left(\partial_{\tau}\Theta\right)^2}{\sinh^2\rho}+\frac{\left(\partial_{\alpha}\Theta\right)^2}{\sin^2\Theta}+\frac{\left(\partial_{\psi}\Theta\right)^2}{\sin^2\Theta\cos^2\alpha}+\frac{F_{\sigma\tau}^2}{\rho'^2\sinh^2\rho}+\frac{F_{\sigma\alpha}^2}{\rho'^2\sin^2\Theta}
	\cr
	&+\frac{F_{\sigma\psi}^2}{\rho'^2\sin^2\Theta\cos^2\alpha}+\frac{F_{\tau\alpha}^2}{\sinh^2\rho\sin^2\Theta}+\frac{F_{\tau\psi}^2}{\sinh^2\rho\sin^2\Theta\cos^2\alpha}+\frac{F_{\alpha\psi}^2}{\sin^4\Theta\cos^2\alpha}
	\cr
	&+\frac{\left(\partial_{\sigma}\Theta\,F_{\tau\alpha}+\partial_{\alpha}\Theta\,F_{\sigma\tau}-\partial_{\tau}\Theta\,F_{\sigma\alpha}\right)^2}{\rho'^2\sinh^2\rho\sin^2\Theta}+\frac{\left(\partial_{\sigma}\Theta\,F_{\tau\psi}+\partial_{\psi}\Theta\,F_{\sigma\tau}-\partial_{\tau}\Theta\,F_{\sigma\psi}\right)^2}{\rho'^2\sinh^2\rho\sin^2\Theta\cos^2\alpha}
	\cr
	&+\frac{\left(\partial_{\sigma}\Theta\,F_{\alpha\psi}+\partial_{\psi}\Theta\,F_{\sigma\alpha}-\partial_{\alpha}\Theta\,F_{\sigma\psi}\right)^2}{\rho'^2\sin^4\Theta\cos^2\alpha}+\frac{\left(\partial_{\tau}\Theta\,F_{\alpha\psi}+\partial_{\psi}\Theta\,F_{\tau\alpha}-\partial_{\alpha}\Theta\,F_{\tau\psi}\right)^2}{\sinh^2\rho\sin^4\Theta\cos^2\alpha}
	\cr
	&+\frac{\left(F_{\sigma\tau}F_{\alpha\psi}-F_{\sigma\alpha}F_{\tau\psi}+F_{\sigma\psi}F_{\tau\alpha}\right)^2}{\rho'^2\sinh^2\rho\sin^4\Theta\cos^2\alpha}\,.
\end{empheq}
Secondly, the D5-brane $\kappa$-symmetry projector \eqref{eq: D5 projector 2} yields (setting $\rho=\sigma$)
\begin{empheq}{alignat=7}\label{eq: D5 projector 3}
	\Gamma_{D5}&=\frac{\Gamma_{\underline{126789}}\sigma_2}{\sqrt{\mathcal{L}}}\left[\left(1+\partial_{\rho}\Theta\Gamma_{\underline{15}}+\frac{\partial_{\tau}\Theta\Gamma_{\underline{25}}}{\sinh\rho}+\frac{\partial_{\alpha}\Theta\Gamma_{\underline{65}}}{\sin\Theta}+\frac{\partial_{\psi}\Theta\Gamma_{\underline{75}}}{\sin\Theta\cos\alpha}\right)\sigma_3\right.
	\cr
	&+\frac{F_{\rho\tau}\Gamma_{\underline{12}}}{\sinh\rho}+\frac{F_{\rho\alpha}\Gamma_{\underline{16}}}{\sin\Theta}+\frac{F_{\rho\psi}\Gamma_{\underline{17}}}{\sin\Theta\cos\alpha}+\frac{F_{\tau\alpha}\Gamma_{\underline{26}}}{\sinh\rho\sin\Theta}+\frac{F_{\tau\psi}\Gamma_{\underline{27}}}{\sinh\rho\sin\Theta\cos\alpha}+\frac{F_{\alpha\psi}\Gamma_{\underline{67}}}{\sin^2\Theta\cos\alpha}
	\cr
	&+\frac{\left(F_{\rho\tau}\partial_{\alpha}\Theta+F_{\tau\alpha}\partial_{\rho}\Theta-F_{\rho\alpha}\partial_{\tau}\Theta\right)\Gamma_{\underline{1265}}}{\sinh\rho\sin\Theta}+\frac{\left(F_{\rho\tau}\partial_{\psi}\Theta+F_{\tau\psi}\partial_{\rho}\Theta-F_{\rho\psi}\partial_{\tau}\Theta\right)\Gamma_{\underline{1275}}}{\sinh\rho\sin\Theta\cos\alpha}
	\cr
	&+\frac{\left(F_{\rho\alpha}\partial_{\psi}\Theta-F_{\rho\psi}\partial_{\alpha}\Theta+F_{\alpha\psi}\partial_{\rho}\Theta\right)\Gamma_{\underline{1675}}}{\sin^2\Theta\cos\alpha}+\frac{\left(F_{\tau\alpha}\partial_{\psi}\Theta-F_{\tau\psi}\partial_{\alpha}\Theta+F_{\alpha\psi}\partial_{\tau}\Theta\right)\Gamma_{\underline{2675}}}{\sinh\rho\sin^2\Theta\cos\alpha}
	\cr
	&\left.+\frac{\left(F_{\rho\tau}F_{\alpha\psi}-F_{\rho\alpha}F_{\tau\psi}+F_{\rho\psi}F_{\tau\alpha}\right)\Gamma_{\underline{1267}}\sigma_3}{\sinh\rho\sin^2\Theta\cos\alpha}\right]\,.
\end{empheq}
Lastly, the expressions involved in the $\frac{1}{4}$-BPS equations \eqref{eq: eq0}-\eqref{eq: eq15} are
\begin{flalign}
   \textrm{eq}_{0}&=\cos\theta_{0}\sin\Theta+\sin\theta_{0}\sin\Theta\cos\alpha\cos\Delta\R-\sin\theta_{0}\coth\rho\sin\Theta\cos\alpha\sin\Delta\T
    \cr
    &+i\sinh\rho\cot\Theta\Frp+i\sinh\rho\csc^2\Theta\tan\alpha\left(\R\Fap+\Ps\Fra-\A\Frp\right)
    \cr
    &-\cos\theta_0\coth\rho\csc\Theta\tan\alpha\left(\Frt\Fap-\Fra\Ftp+\Frp\Fta\right)\,,
    \\
    \textrm{eq}_{1}&=-\sin\theta_0\sin\Theta\cos\alpha\sin\Delta\R-\left(\cos\theta_0\cos\Theta+\sin\theta_0\coth\rho\sin\Theta\cos\alpha\cos\Delta\right)\T
    \cr
    &-i\,\textrm{csch}\,\rho\left(\R\Ftp+\Ps\Frt-\T\Frp\right)\,,
    \\
    \textrm{eq}_{2}&=-\cos\theta_0\sin\Theta\sin\alpha\cos\Delta\R+\cos\theta_0\coth\rho\sin\Theta\sin\alpha\sin\Delta\T
    \cr
    &+i\cosh\rho\cos\alpha\sin\Delta\Fra+i\,\textrm{csch}\,\rho\cos\alpha\cos\Delta\Fta+i\sinh\rho\csc^2\Theta\sec\alpha\cos\Delta\Fap
    \cr
    &+\csc\Theta\left(\sin\theta_0\coth\rho+\cos\theta_0\cot\Theta\sec\alpha\cos\Delta\right)\left(\Frt\Fap-\Fra\Ftp+\Frp\Fta\right)\,,
    \\
    \textrm{eq}_{3}&=\cos\theta_0\sin\Theta\cos\alpha\sin\Delta\R-\left(\sin\theta_0\cos\Theta-\cos\theta_0\coth\rho\sin\Theta\cos\alpha\cos\Delta\right)\T
    \cr
    &-i\cosh\rho\sin\alpha\cos\Delta\Fra+i\cosh\rho\sec\alpha\sin\Delta\Frp+i\,\textrm{csch}\,\rho\sin\alpha\sin\Delta\Fta
    \cr
    &+i\,\textrm{csch}\,\rho\sec\alpha\cos\Delta\Ftp\,,
    \\
    \textrm{eq}_{4}&=-\cos\theta_0\sin\Theta\sin\alpha\sin\Delta\R-\cos\theta_0\coth\rho\sin\Theta\sin\alpha\cos\Delta\T
    \cr
    &-i\cosh\rho\cos\alpha\cos\Delta\Fra+i\,\textrm{csch}\,\rho\cos\alpha\sin\Delta\Fta+i\sinh\rho\csc^2\Theta\sec\alpha\sin\Delta\Fap
    \cr
    &+\cos\theta_0\cot\Theta\csc\Theta\sec\alpha\sin\Delta\left(\Frt\Fap-\Fra\Ftp+\Frp\Fta\right)\,,
    \\
    \textrm{eq}_{5}&=\sin\theta_0\sin\Theta-\cos\theta_0\sin\Theta\cos\alpha\cos\Delta\R+\cos\theta_0\coth\rho\sin\Theta\cos\alpha\sin\Delta\T
    \cr
    &-i\cosh\rho\sin\alpha\sin\Delta\Fta-i\cosh\rho\sec\alpha\cos\Delta\Frp
    \cr
    &-\sin\theta_0\coth\rho\csc\Theta\tan\alpha(\Frt\Fap-\Fra\Ftp+\Frp\Fta)
    \cr
    &-i\,\textrm{csch}\,\rho\sin\alpha\cos\Delta\Fta+i\,\textrm{csch}\,\rho\sec\alpha\sin\Delta\Ftp,
    \\
    \textrm{eq}_{6}&=\sin\theta_0\sin\Theta\sin\alpha\sin\Delta\R+\sin\theta_0\coth\rho\sin\Theta\sin\alpha\cos\Delta\T
    \cr
    &-\sin\theta_0\csc\Theta\cot\Theta\sec\alpha\sin\Delta\left(\Frt\Fap-\Fra\Ftp+\Frp\Fta\right)-i\sinh\rho\cot\Theta\Fra
    \cr
    &+i\,\textrm{csch}\,\rho\tan\alpha\left(\R\Ftp-\T\Frp+\Ps\Frt\right)\,,
    \\
    \textrm{eq}_{7}&=-\cos\theta_0\cos\Theta\sin\Delta\R-\left(\cos\theta_0\coth\rho\cos\Theta\cos\Delta+\sin\theta_0\sin\Theta\cos\alpha\right)\T
    \cr
    &-\sin\theta_0\csc\Theta\sec\alpha\Ps-i\cosh\rho\cot\Theta\sin\Delta\Frp-i\,\textrm{csch}\,\rho\cot\Theta\cos\Delta\Ftp
    \cr
    &+i\cosh\rho\csc^2\Theta\tan\alpha\sin\Delta\left(\R\Fap+\Ps\Fra-\A\Frp\right)
    \cr
    &+i\,\textrm{csch}\,\rho\csc^2\Theta\tan\alpha\cos\Delta(\T\Fap+\Ps\Fta-\A\Ftp)
    \cr
    &-\cos\theta_0\csc\Theta\tan\alpha\sin\Delta\left(\Frt\Fap-\Fra\Ftp+\Frp\Fta\right)\,,
    \\
    \textrm{eq}_{8}&=-\sin\theta_0\cos\Theta\cos\alpha+\cos\theta_0\cos\Theta\cos\Delta\R-\cos\theta_0\coth\rho\cos\Theta\sin\Delta\T
    \cr
    &-\sin\theta_0\csc\Theta\sin\alpha\A+i\cosh\rho\cot\Theta\cos\Delta\Frp-i\,\textrm{csch}\,\rho\cot\Theta\sin\Delta\Ftp
    \cr
    &-i\cosh\rho\csc^2\Theta\tan\alpha\cos\Delta\left(\R\Fap+\Ps\Fra-\A\Frp\right)
    \cr
    &+i\,\textrm{csch}\,\rho\csc^2\Theta\tan\alpha\sin\Delta\left(\T\Fap+\Ps\Fta-\A\Ftp\right)
    \cr
    &+\cos\theta_0\csc\Theta\tan\alpha\cos\Delta\left(\Frt\Fap-\Fra\Ftp+\Frp\Fta\right)\,,
    \\
    \textrm{eq}_{9}&=\cos\theta_0\sin\Theta\sin\alpha\T+i\sinh\rho\cos\alpha\Fra
    \cr
    &+\sin\theta_0\csc\Theta\sin\Delta\left(\Frt\Fap-\Fra\Ftp+\Frp\Fta\right)\,,
    \\
    \textrm{eq}_{10}&=\cos\theta_0\cos\Theta\cos\alpha+\sin\theta_0\cos\Theta\cos\Delta\R-\sin\theta_0\coth\rho\cos\Theta\sin\Delta\T
    \cr
    &+\sin\theta_0\csc\Theta\tan\alpha\cos\Delta\left(\Frt\Fap-\Fra\Ftp+\Frp\Fta\right)+\cos\theta_0\csc\Theta\sin\alpha\A
    \cr
    &-i\,\textrm{csch}\,\rho\cot\Theta\sin\alpha\left(\R\Fta+\A\Frt-\T\Fra\right)-i\,\textrm{csch}\,\rho\cos\alpha\Frt
    \cr
    &+i\sinh\rho\sec\alpha\Frp\,,
    \\
    \textrm{eq}_{11}&=\sin\theta_0\sin\Theta\sin\alpha\cos\Delta\R-\sin\theta_0\coth\rho\sin\Theta\sin\alpha\sin\Delta\T
    \cr
    &-i\sinh\rho\cot\Theta\tan\alpha\Frp-i\,\textrm{csch}\,\rho\left(\R\Fta+\A\Frt-\T\Fra\right)
    \cr
    &-i\sinh\rho\csc^2\Theta\left(\R\Fap+\Ps\Fra-\A\Frp\right)
    \cr
    &+\csc\Theta\left(\cos\theta_0\coth\rho-\sin\theta_0\cot\Theta\sec\alpha\cos\Delta\right)\left(\Frt\Fap-\Fra\Ftp+\Frp\Fta\right)\,,
    \\
    \textrm{eq}_{12}&=-\cos\theta_0\cos\Theta\sin\alpha+\cos\theta_0\csc\Theta\cos\alpha\A+i\,\textrm{csch}\,\rho\sin\alpha\Frt
    \cr
    &+i\cosh\rho\csc^2\Theta\sec\alpha\Fap-i\,\textrm{csch}\,\rho\cot\Theta\cos\alpha\left(\R\Fta+\A\Frt-\T\Fra\right)
    \cr
    &+\csc\Theta\left(\sin\theta_0\cos\Delta+\cos\theta_0\coth\rho\cot\Theta\sec\alpha\right)\left(\Frt\Fap-\Fra\Ftp+\Frp\Fta\right)\,,
    \\
    \textrm{eq}_{13}&=-\sin\theta_0\cos\Theta\sin\Delta\R+\left(\cos\theta_0\sin\Theta\cos\alpha-\sin\theta_0\coth\rho\cos\Theta\cos\Delta\right)\T
    \cr
    &-\sin\theta_0\csc\Theta\tan\alpha\sin\Delta\left(\Frt\Fap-\Fra\Ftp+\Frp\Fta\right)+\cos\theta_0\csc\Theta\sec\alpha\Ps
    \cr
    &-i\,\textrm{csch}\,\rho\cot\Theta\sec\alpha\left(\R\Ftp+\Ps\Frt-\T\Frp\right)-i\sinh\rho\sin\alpha\Fra\,,
    \\
    \textrm{eq}_{14}&=\sin\theta_0\cos\Theta\sin\alpha-\sin\theta_0\csc\Theta\cos\alpha\A
    \cr
    &-i\cosh\rho\cot\Theta\sin\Delta\Fra-i\cosh\rho\cot\Theta\tan\alpha\cos\Delta\Frp-i\,\textrm{csch}\,\rho\cot\Theta\cos\Delta\Fta
    \cr
    &+i\,\textrm{csch}\,\rho\cot\Theta\tan\alpha\sin\Delta\Ftp-i\cosh\rho\csc^2\Theta\cos\Delta\left(\R\Fap+\Ps\Fra-\A\Frp\right)
    \cr
    &+i\,\textrm{csch}\,\rho\csc^2\Theta\sin\Delta\left(\T\Fap+\Ps\Fta-\A\Ftp\right)
    \cr
    &+\csc\Theta\left(\cos\theta_0\cos\Delta-\sin\theta_0\coth\rho\cot\Theta\sec\alpha\right)\left(\Frt\Fap-\Fra\Ftp+\Frp\Fta\right)\,,
    \\
    \textrm{eq}_{15}&=-\sin\theta_0\sin\Theta\sin\alpha\T+i\cosh\rho\cot\Theta\cos\Delta\Fra-i\cosh\rho\cot\Theta\tan\alpha\sin\Delta\Frp
    \cr
    &-i\,\textrm{csch}\,\rho\cot\Theta\sin\Delta\Fta-i\,\textrm{csch}\,\rho\cot\Theta\tan\alpha\cos\Delta\Ftp
    \cr
    &-i\cosh\rho\csc^2\Theta\sin\Delta\left(\R\Fap+\Ps\Fra-\A\Frp\right)
    \cr
    &-i\,\textrm{csch}\,\rho\csc^2\Theta\cos\Delta\left(\T\Fap+\Ps\Fta-\A\Ftp\right)
    \cr
    &+\cos\theta_0\csc\Theta\sin\Delta\left(\Frt\Fap-\Fra\Ftp+\Frp\Fta\right)\,.
\end{flalign}
\bibliographystyle{JHEP}
\bibliography{D5}
\end{document}